\documentclass[12pt]{article}

\usepackage{amsmath}
\usepackage{graphicx}
\newcounter{eqncnt}[section]

\def\SU2{\ensuremath{{SU(2)}}}

\newcommand{\bea}{\begin{eqnarray}}
\newcommand{\eea}{\end{eqnarray}}

\def\beq{\begin{equation}}
\def\eeq{\end{equation}}
\def\beqn{\begin{eqnarray}}
\def\eeqn{\end{eqnarray}}

\begin{document}
\begin{center}
{\bf Persistence of black holes through a cosmological bounce } \\[2mm]

\vskip .5in

{\sc B.~J.~Carr}\\
{\it  Astronomy Unit, Queen Mary University of London}\\
{\it Mile End Road, London E1 4NS, UK}\\
{\it B.J.Carr@qmul.ac.uk }

\vskip .5in

{\sc A.~A.~Coley}\\
{\it Department of Mathematics and Statistics}\\
{\it Dalhousie University, Halifax, NS B3H 3J5, Canada}\\
{\it aac@mathstat.dal.ca }

\end{center}

\date{\today}

\begin{abstract}

We discuss whether black holes could 
persist in a universe
which recollapses and then bounces into a new expansion phase. 
Whether the bounce is of classical or
quantum gravitational origin, such cosmological models are of great current interest.
In particular, we investigate the mass range in which black holes might survive a
bounce and ways of
differentiating observationally
between black holes formed just after and just before the last
bounce. We also discuss the consequences of the universe going
through a sequence of dimensional changes as it passes through
a bounce.

\end{abstract}

Essay written for the Gravity Research Foundation 2011 Awards for Essays on Gravitation; submission date  March 28, 2011

\newpage


In some cosmological scenarios, the universe is expected to recollapse to a big crunch in the future and then bounce into a new expansion phase. 
The evidence that the universe is currently accelerating does not exclude this possibility 
if the acceleration is driven by a scalar field rather a 
cosmological constant \cite{caldwell}.
Even if the universe is destined to expand forever, its present expanding phase may have been preceded by the collapse and bounce of an earlier universe. 
Both past and future bounces would arise in cyclic models but not all bouncing models are cyclic.

As regards the bounce mechanism,
even classical general relativity (GR) permits a 
turn-around if one invokes a positive cosmological constant \cite{lemaitre}, 
although this simple option is not favoured by current observations. Other possible mechanisms are
based on
alternative theories of gravity (such as higher derivative theories \cite{novello}) which lead to
a modified Friedmann equation. 
A bounce can also occur within theories of quantum gravity, such as
string theory \cite{string}, where it 
has been dubbed the ``big bounce" within the {\em pre-big-bang} scenario \cite{BHScorr}, loop quantum gravity (LQG), where  
both black hole and cosmological 
singularities may be resolved \cite{Sing}, or quantum cosmology \cite{hartle}. 
The bounce density would most naturally have the Planck value in quantum gravity but it could be
sub-Planckian in some string 
scenarios or if one invokes a classical mechanism.

Cyclic models were included even in the original Friedmann paper \cite{friedmann}, although without specifying any bounce mechanism.  A more physical GR model was proposed by Tolman  
\cite{tolman}, in which extra entropy is generated at each bounce, leading to universes which attain progressively larger size at maximum expansion. LQG versions of this type of model might allow the universe to eventually escape the cyclic phase and enter a final de Sitter period \cite{tavakol}. Another interesting scenario is {\em cyclic brane cosmology}
\cite{cyclic},  in which the universe undergoes a periodic (and 
classically eternal) 
sequence of big bangs and big crunches, each one being associated with
the collision of two 3-branes in a fourth spatial dimension. 
Each bang results in
baryogenesis, dark matter production,
nucleosynthesis, galaxy formation, 
an epoch of low-energy acceleration, 
and finally a contraction that produces homogeneity and flatness in the next cycle.
There is no inflation but the dynamical behavior in the final phase of
each cycle is supposed to explain many of the observational features usually attributed to inflation. 
The nature of the fluctuations generated by quantum perturbations 
in this model
has been studied in \cite{perts}. 
\if
In {\em conformal cyclic cosmology} (CCC) \cite{penrose} there exists
an unending succession of
aeons and our big bang is identified with the conformal infinity of the previous one.
In the standard interpretation, CCC does not involve a collapsing phase or bounce, but comparison with the ``pre-big-bang" model of string theory suggests that this interpretation might conceivably apply in a different frame.
\fi

Whatever the scenario, black holes would be an important probe of a cosmological bounce,
just as primordial black holes (PBHs) 
provide an important probe of the early stages of the standard big bang~\cite{cksy}.  
However, the issues raised are somewhat different for future and past bounces. 
Both scenarios raise the
question of whether black holes formed in one universe can persist into the next but the existence of a past bounce directly impinges on observations. 
In particular, we 
need to distinguish between black holes which formed {\it during} the big crunch
(i.e.  because of it) and those that formed {\it before} it.  We refer to these as
``big-crunch black holes'' (BCBHs) and ``pre-crunch black holes'' (PCBHs), respectively. It is not clear whether 
any observations could distinguish between black holes which form just before the bounce (i.e., in the final moments of the big crunch) or just afterwards (i.e., in the first moments of the big bang), since both types of black holes would have comparable mass ranges.

Future bounces are obviously not relevant to current observations but they raise the mathematical issue of what is meant by a black hole in
recollapsing universes.  Since these have closed spatial hypersurfaces,
there is no asymptotic spatial infinity, so  the whole universe is in a sense a black
hole and each black hole singularity is part of the future big
crunch one. By contrast, the black hole singularity in an ever-expanding model is not part of any cosmological singularity. Of course, all
singularities may be removed in a bouncing model. 

\if
The production of primordial black hole (PBH) remnants in the
early (pre-inflationary) universe  was considered in  \cite{ACSantiago}.
It was found that the
effects of such an
epoch might explain the observed WMAP quadrupole anomaly in the CMB power spectrum.
A salient feature of
this BH nucleation is that the production rate per unit volume
is very high at the Planck temperature.   However,
the complete decay of the nucleated BH into radiation is
prevented, and we have massive but inert black hole remnants
\cite{ACSantiago} populating the pre-inflationary phase of the universe.
\fi

\begin{figure}
 \begin{center}
 \includegraphics[height=7cm]{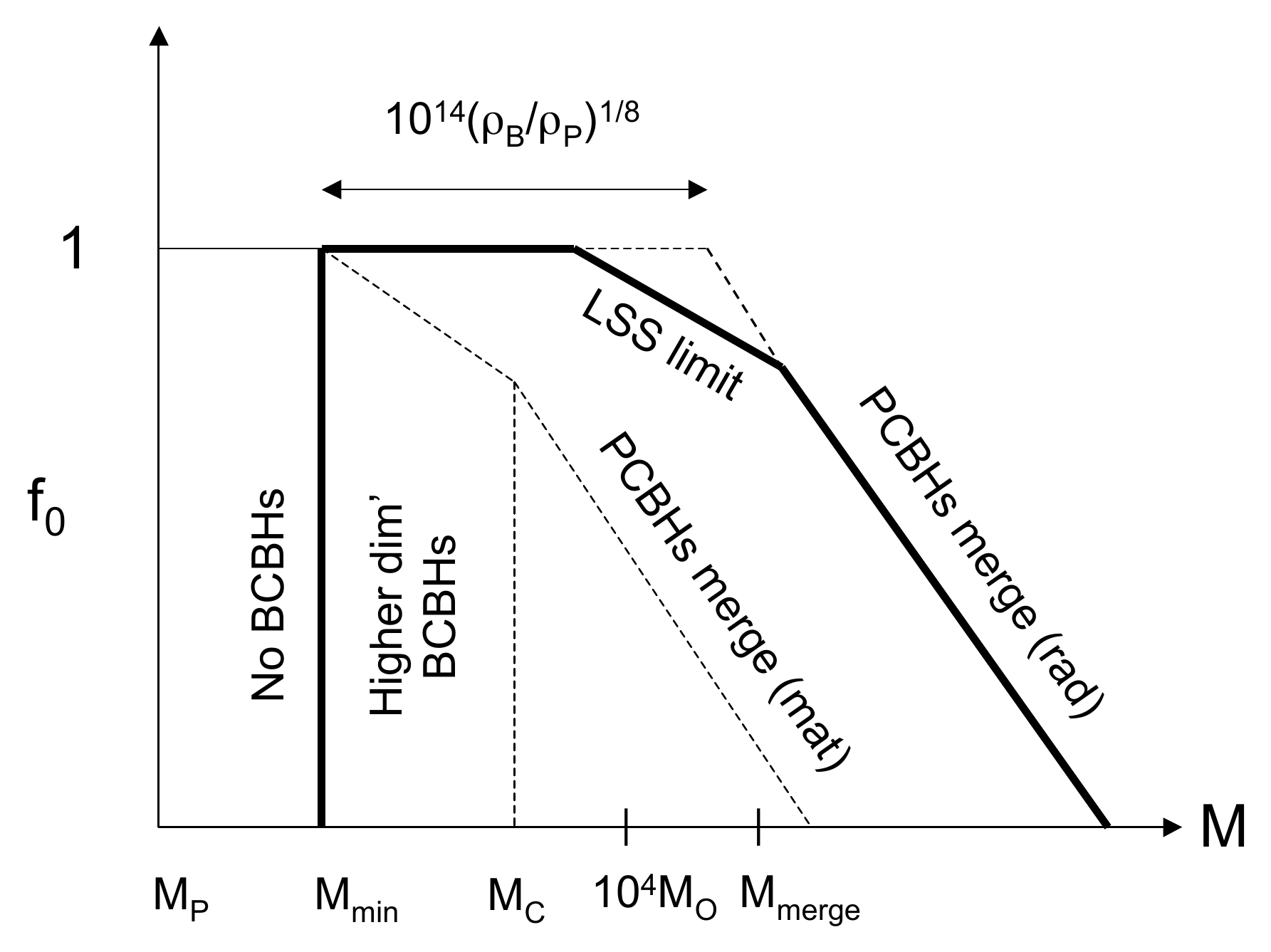}
   \end{center}
  \caption{ \label{bounce} This shows the domain in which black holes of mass $M$ containing a fraction $f_0$ of the present density can form in the big crunch or avoid merging if they exist before then. }
  \end{figure}

Let us first consider the mass range in which black holes can 
form in a bounce.
We assume that the universe bounces at some density $\rho_B$, which 
might either be of order the Planck density,
$\rho_P  \sim c^5/(G^2  \hbar) \sim 10^{94}\mathrm{g \, cm}^{-3}$, or much less. 
A spherical region of mass $M$ becomes a black hole when it falls within 
its Schwarzschild radius, $R_{S}= 2GM/c^2$.
At this point the collapsing matter has density  
$\rho_{BH} = (3M/4\pi R_{S}^3) 
\sim 10^{18}({M}/{M_{\odot}})^{-2} \mathrm{g \, cm}^{-3} $
from the perspective of an external observer, although the matter itself collapses to a greater density.  
A BCBH can presumably only form if its density is less than $\rho_B$, and
this corresponds to a {\it lower} limit on the black hole mass $M_{\mathrm{min}} \sim (\rho_P/\rho_B)^{1/2}M_P$,
where $M_P \sim 10^{-5}$g is the Planck mass. This corresponds to the bold line on the left of Fig.~\ref{bounce}. BCBH formation is not guaranteed by the density alone, since one would expect it to require inhomogeneities or some form of phase transition. However, it is possible that the
black hole nucleation rate might become
very high at the Planck temperature for quantum gravitational reasons. 

There is also a mass range in which pre-existing PCBHs 
lose their individual identity by merging with each other prior to the bounce.  
If the fraction of the cosmological density in these black holes at the 
bounce epoch is $f_B$, 
then the average separation between them is less than their size (i.e. the black holes merge) for 
$M > f_B^{-1/2} M_{\mathrm{min}} $.
This condition can also be written as
  $f_B > (M/M_{\mathrm{min}})^{-2} $. 
Since $f_B$ cannot exceed $1$, there is a always range of masses in which BCBHs may form and PCBHs do not merge. If the PCBHs do merge,  this will initially  generate holes with a hierarchy of masses larger than $M$ but eventually the universe will be converted into a homogenous vacuum state apart from the sprinkling of singularities (or ``Planck balls'') generated by the original collapsing matter. 
\vspace*{0.5cm}

Unless the universe is always matter-dominated, one must distinguish between $f_B$ and the {\it present} fraction $f_0$ of the universe's mass in black holes.   
The ratio of the matter to radiation density 
scales as the cosmic scale factor $R$, so this will decrease during collapse and increase during expansion in a radiation-dominated era. 
If $\rho_0$ and $\rho_{rad}$ are the {\it current} densities 
of the matter (including black holes) and cosmic background radiation, matter-radiation equality therefore occurs when
$R_{eq}/R_0 = \rho_{rad}/ \rho_0 \sim 10^{-4}$. This corresponds to a density $\rho_{\mathrm{eq}} \sim 10^{12} \rho_0 \sim 10^{-17} \mathrm{g \, cm}^{-3}$ and this applies in either a contracting or
expanding phase. The fraction of the universe in black holes 
at a radiation-dominated bounce is therefore
$f_B 
 \approx  
f_0 \left({\rho_\mathrm{eq}}/{\rho_B} \right) ^{1/4}$, so
the merger condition becomes $f_0 > \left({M/ M_{\mathrm{min}} } \right)^{-2} \left({\rho_\mathrm{B}}/{\rho_{\mathrm{eq}}} \right) ^{1/4}$.
Substituting for $M_{\mathrm{min}}$ and $\rho_{\mathrm{eq}}$ gives $f_0 > 10^{28}\left({\rho_\mathrm{B}}/{\rho_{\mathrm{P}}} \right) ^{-3/4}\left(M/M_P \right)^{-2} $, 
corresponding to the bold line on the right of Fig.~\ref{bounce}. 
Equivalently, 
 $M > 10^{14} f_0^{-1/2} (\rho_P/ \rho_B) ^{3/8} M_P$.

There are various dynamical 
constraints on the form of the function $f_0(M)$ for PCBHs which are non-evaporating (i.e., larger than $M_* \sim 10^{15}$g) \cite{cksy}.
The most obvious one is that 
non-evaporating black holes must have $f_0 < 1$ in order not to exceed the
observed cosmological density and this gives a minimum  value for the merger mass.
We can express this as $M_{\mathrm{merge}} \sim 10^{9} (t_B/ t_P) ^{3/4}$~g, where $t_B$ is the time of the bounce
as measured from the notional time of infinite density. 
This is around $10^{15}$~g for $t_B \sim 10^{-35}$~s but as large as $1M_{\odot}$ for 
$t_B \sim 
10^{-13}$s or 
$10^4 M_{\odot}$ for $t_B \sim 10^{-5}$s, so the observational consequences would be very significant. 
Note that the density of the universe when the black holes merge is much greater than it is when they form. Another important dynamical constraint is associated with
large-scale structure (LSS) formation deriving from Poisson fluctuations in the
black hole number density
\cite{Meszaros}. Applying this argument to the formation of Lyman-$\alpha$ clouds  gives $f_0 < (M/10^4 M_{\odot})^{-1}$~\cite{afshordi}, implying an
upper limit of $10^4M_\odot $ on the mass of any black holes which provide
the dark matter.
These limits are shown by the bold lines at the top of Fig.~\ref{bounce}. 
We note that in some quantum gravity models, the complete decay of black holes is
prevented, leading to stable remnants with around the Planck mass. This leads to a limit 
$f_0 < (M/M_P)$ over some range of $M$ but this is not shown explicitly in the figure.

One important observational issue is whether it is possible to differentiate
between black holes formed just before and after the bounce. In the standard non-bouncing scenario, PBHs generated {\it before} inflation are exponentially diluted, so PBHs present today are usually assumed to form at the {\it end} of inflation. In a bouncing model, unless there were a  matching deflationary period in the collapse phase, any BCBHs would also be exponentially diluted.  On the other hand, there is no inflation in the expanding phase of some bouncing  models. Another difference concerns the relative importance of accretion and evaporation. Accretion is negligible in the expanding phase but may not be in the collapsing phase and this could block evaporation altogether. If evaporation does occur, it would be on a much longer timescale than the black hole formation time and so would occur after the bounce, whenever the black holes form.

It is possible that the universe gains an extra spatial dimension -- or goes
through a sequence of dimensional increases -- as it approaches the big bang in the past or the big
crunch in the future.  
The scale of the extra dimension, $R_C$, would most naturally be comparable to the Planck
length but it could be much larger, and the time of the transition (measured notionally from  the time of infinite density) 
would be $\tau_C \sim R_C/c$. The associated density is then
$\rho_C \sim 
c^2/(G R_C^2)$, which
is less than the bounce density (so that higher-dimensional effects are important) for
$R_C  > c (G \rho_B)^{-1/2}$.
At larger densities both the matter content and any black holes must be regarded as higher dimensional. 

If there are $3+n$ spatial dimensions and the black hole is spherically symmetric in all of them,
 then the radius of its event horizon becomes
\if
$R_S  \sim  \left({M}/{M_P'} \right) ^{1/(n+1)} R_P'$, where $M_P'$ and $R_P'$ 
are the revised
Planck scales, {\bf which will be different from the standard values if there are large extra dimensions.}
\fi
$R_S   
 \sim  \left({M}/{M_C} \right) ^{1/(n+1)} R_C$,
where $M_C \sim c^2 R_C/(2G)$ is the mass at which the dimensional transition occurs.
Thus $R_S \propto M$ for $n=0$ 
and $R_S \propto M^{1/2}$ for $n=1$.
\if
For $R_C \gg R_P$, the revised Planck scales are 
$R_P' \sim R_P \left({R_C}/{R_P} \right) ^{n/(n+2)} \gg M_P $ and   
$M_P' \sim M_P \left({R_P}/{R_C} \right) ^{n/(n+2)} \ll R_P$.
We then have
\fi
For $M <M_C$, the higher-dimensional density required for black hole formation is
$\rho_{BH} \propto  M^{-2/(1+n)}$,
so the lower limit on the BCBH mass becomes
\if
$M_{\mathrm{min}} \sim 
\left(\rho_P'/ \rho_B \right)^{(1+n)/2} M_P' $
where both $\rho_{BH}$ and $\rho_P$ are 
now interpreted as higher dimensional densities. The lower limit can also be written as
\fi 
$M_{\mathrm{min}} \sim 
\left({\rho_P}/{\rho_B} \right)^{(1+n)/2} \left({R_P}/{R_C} \right) ^{n(3+n)/2}  M_P$.
The black hole merger condition
becomes
$f_B > \left({M}/{M_
{\mathrm{min}}} \right)^{-2/(1+n)}$,
which is further modified if the universe is dominated by its higher-dimensional matter content when it bounces.
Conceivably, the change in dimensionality could itself trigger a bounce. 
\if
in which case $\rho_B/\rho_P \sim (R_P/R_C)^{3+n}$. We then obtain
$M_{\mathrm{min}} \sim 
\left({R_C}/{R_P} \right) ^{(3+n)/2}  M_P$,
which is necessarily larger than before.
\fi

To conclude, we have discussed whether black holes in some mass range could persist in a
universe which recollapses and then bounces into a new expansion phase.
We find that there is a range of masses in which BCBHs form and 
PCBHs do not merge
but these limits are
modified by the inclusion of radiation and the effects of
extra dimensions.
The consequences of such black holes, only some of which have been discussed here, provide an important signature of bouncing cosmologies, which allows them to
be falsified by observations. 
One problem which has stymied the
success of cyclic models is that the formation of large-scale
structure and black holes during the expanding phase leads to
difficulties during the contracting phase \cite{tolman}. This may be related to the second law 
of thermodynamics because unless the black
holes are small enough to evaporate via Hawking
radiation, the area theorems imply that they
grow ever larger during subsequent cycles. Understanding the persistence of black holes through a bounce is clearly relevant to this problem. 

Finally, some of these arguments may also pertain in cosmological scenarios which do not involve a bounce. For example, in the {\em cosmological natural selection} scenario, black holes evolve into separate expanding universes 
and this means that 
the constants of physics evolve so as to maximize the
number of black holes \cite{LOTC}. This raises the issue of whether the persistence of black holes is related to
the second law of thermodynamics. In the {\em conformal cyclic cosmology},
there exists an unending succession of aeons and our
big bang is identified with the conformal infinity of the previous
one \cite{penrose}.  Numerous supermassive black hole encounters occurring within
clusters of galaxies in the previous aeon would yield bursts of
gravitational radiation and these would generate randomly distributed
families of concentric circles in the CMB sky over which the
temperature variance is anomalously low \cite{penrose}.

\if
Finally, some of these arguments may also pertain in cosmological scenarios which do not involve a bounce. For example, in the {\em conformal cyclic cosmology} \cite{penrose} there exists
an unending succession of
aeons and our big bang is identified with the conformal infinity of the previous one.
Numerous supermassive
black hole encounters occurring within clusters of galaxies in the previous aeon
would yield bursts of
gravitational radiation and these would generate randomly distributed families of concentric circles  in the CMB sky over which the
temperature variance is anomalously low  \cite{penrose}.
Also relevant is the {\em cosmological natural selection} scenario, in which -- rather than the whole universe bouncing -- the matter collapsing into a black hole evolves into another expanding universe  \cite{LOTC}. In this scenario the constants of physics evolve so as to maximize the number of black holes.
\fi
\if
This is different from the whole universe bouncing but the ideas may be related since a collapsing universe is in a sense a black hole.
In LQG for small enough masses no (dynamical) horizon develops (i.e. there is no black hole),
so the bounce is naked~\cite{Sing}, which suggests that 
quantum gravity may exclude very small
astrophysical black holes. 
\fi



\begin{thebibliography}{99}

\bibitem{caldwell}
R. Caldwell, Phys, Lett. B. {\bf 545}, 23 (2002).

\bibitem{lemaitre}
G. Lemaitre, Mon. Not. R. Astron. Soc. {\bf 91}, 490 (1931)


\bibitem{novello} M. Novello and S. Bergliaffa,
Phys. Reports. {\bf 463}, 127 (2008).



\bibitem{string} 
 N. Turok, M. Perry and P. Steinhardt, Phys. Rev. D {\bf{70}}, 106004 (2004).
 
\bibitem{BHScorr}  
G. Veneziano, Europhys. Lett. {\bf 2}, 133 (1986).

\bibitem{Sing}
M.\ Bojowald, Phys.\ Rev.\ Lett. {\bf 86}, 5227 (2001)
 \& Phys.\ Rev.\ Lett. {\bf 95}, 091302 (2005);
A.~Ashtekar, T.~Pawlowski and P.~Singh,
  Phys.\ Rev.\  D {\bf 74}, 084003 (2006).

\bibitem{hartle}
J.  B. Hartle and S. W. Hawking, Phys. Rev. D. {\bf 28}, 2960 (1983).

\bibitem{friedmann}
A. Friedmann, Zeitschrift fŸr Physik {\bf 10}, 377Ð386 (1922). 

\bibitem{tolman} R. Tolman, {\it Relativity, Thermodynamics, 
and Cosmology}
(Oxford, Clarendon, 1934).

\bibitem{tavakol}
D. J. Mulryne {\it et al.}, Phys. Rev. D. {\bf 71}, 123512 (2005).

\bibitem{cyclic} P.~J.~Steinhardt and N.~Turok,
Phys. Rev. D {\bf 65}, 126003 (2002); J. Khoury, 
P. Steinhardt, and N. Turok, Phys. Rev. Lett. {\bf 92}, 
031302 (2004).
    

\bibitem{perts} J. L. Lehners {\em{et al.}}, Phys. Rev D {\bf 76}, 103501 (2007); Y. Cai {\em{et al.}}
[arXiv:1101.0822].

\bibitem{cksy}
B. J. Carr, K. Kohri, Y. Sendouda and J. Yokoyama, Phys. Rev. D.  {\bf 81}, 104019  (2010).


\bibitem{Meszaros} 
P.Meszaros, Astron. Astrophys.{\bf 38}, 5 (1975); 
B.J.Carr, Astron. Astrophys.{\bf 56}, 377 (77).

\bibitem{afshordi}
N. Afshordi {\em{et al.}}, Ap. J. {\bf 594}, L71 (2003).







\bibitem{LOTC} L. Smolin,    
Class. Q. Grav.  {\bf 9},  173 (1992).



\bibitem{penrose} R.  Penrose, \emph{Cycles of time:  An extraordinary
new view of the universe} (Bodley Head, London, 2010); V.  G.  
Gurzadyan and R.  Penrose, arXiv:1011.3706.

\end{thebibliography}
\end{document}